\documentclass{article}
\usepackage{spconf,amsmath,graphicx,multirow}

\usepackage{enumitem}
\setlist{nosep, leftmargin=14pt}

\usepackage{mwe} 

\begin{document} 
\title{ASIST: Annotation-free synthetic instance segmentation and tracking for microscope video analysis}

\name{\begin{tabular}{c}
Quan Liu $^{\star}$ \quad Isabella M. Gaeta $^{\dagger}$ \quad Mengyang Zhao $^{\mathsection}$ \quad Ruining Deng $^{\star}$ \quad Aadarsh Jha $^{\star}$ 
\\ \textit{Bryan A. Millis} $^{\dagger}$ \quad \textit{Anita Mahadevan-Jansen} $^{\mathparagraph}$ \quad \textit{Matthew J. Tyska} $^{\dagger}$ \quad \textit{Yuankai Huo} $^{\star}$ \end{tabular}}

\address{$^{\star}$ Vanderbilt University, Computer Science, Nashville, TN, USA 37215  \\
$^{\dagger}$ Vanderbilt University, Cell and Developmental Biology, Nashville, TN, USA 37215 \\
$^{\mathsection}$ Tufts University, Computer Science, Medford, MA, USA 02155 \\
$^{\mathparagraph}$ Vanderbilt University, Biomedical Engineering, Nashville, TN, USA 37215\\
}

\maketitle

\begin{abstract}
Instance object segmentation and tracking provide comprehensive quantification of objects across microscope videos. The recent single-stage pixel-embedding based deep learning approach has shown its superior performance compared with "segment-then-associate" two-stage solutions. However, one major limitation of applying a supervised pixel-embedding based method to microscope videos is the resource-intensive manual labeling, which involves tracing hundreds of overlapped objects with their temporal associations across video frames. Inspired by the recent generative adversarial network (GAN) based annotation-free image segmentation, we propose a novel annotation-free synthetic instance segmentation and tracking (ASIST) algorithm for analyzing microscope videos of sub-cellular microvilli. The contributions of this paper are three-fold: (1) proposing a new annotation-free video analysis paradigm is proposed. (2) aggregating the embedding based instance segmentation and tracking with annotation-free synthetic learning as a holistic framework; and (3) to the best of our knowledge, this is first study to investigate microvilli instance segmentation and tracking using embedding based deep learning. From the experimental results, the proposed annotation-free method achieved superior performance compared with supervised learning.
\end{abstract}

\section{Introduction}
\label{sec:intro}

Instance segmentation and tracking provide comprehensive spatial-temporal measurements of objects, which play a critical role in microscopic image analysis. For instance, dense instance segmentation and tracking can be applied on microvilli actin bundle-supported surface protrusions that play a critical role in diverse epithelial cell functions~\cite{Gaeta2020.10.21.341248}. The instance level segmentation and tracking of microvilli provide a quantitative means to understand the formation of the brush border~\cite{meenderink2019actin}. Recent pixel-embedding based supervised deep learning approaches~\cite{zhao2020faster,payer2018instance} have shown superior performance compared with traditional methods ~\cite{al2018deep,korfhage2020detection,van2016deep}, which is a promising solution for segmenting dense and highly overlapped microvilli. However, the deployment of supervised learning is limited by preparing high-quality training data with resource-intensive instance level manual annotation across all videos.

\begin{figure}
\begin{center}
\includegraphics[width=0.9\linewidth]{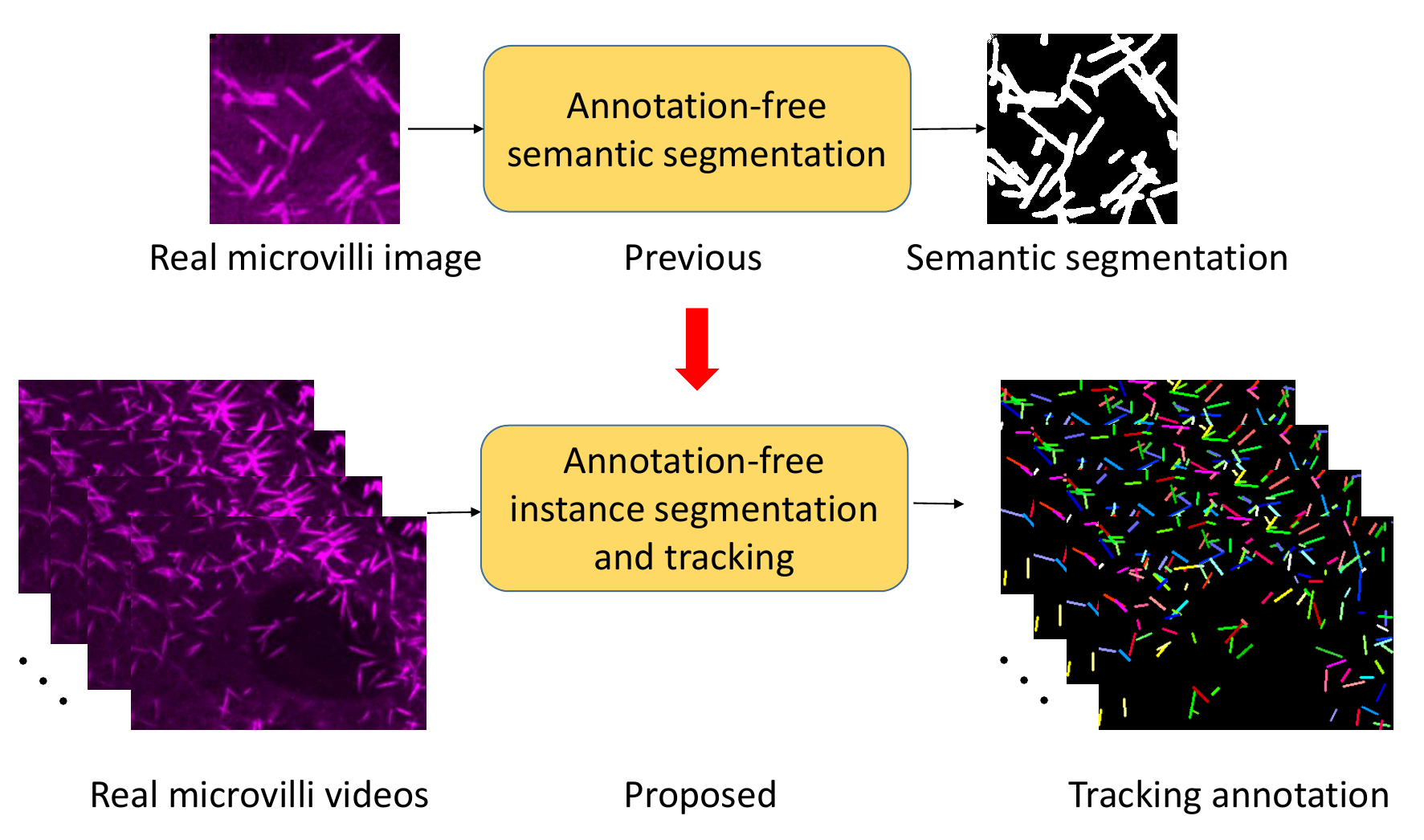}
\end{center}
\caption{As opposed to the previous annotation-free image segmentation method, the proposed ASIST method is able to perform annotation-free video quantification with instance microvilli segmentation and tracking.}
\label{fig:fig1}
\end{figure}


\begin{figure*}[h]
\begin{center}
\includegraphics[width=0.8\linewidth]{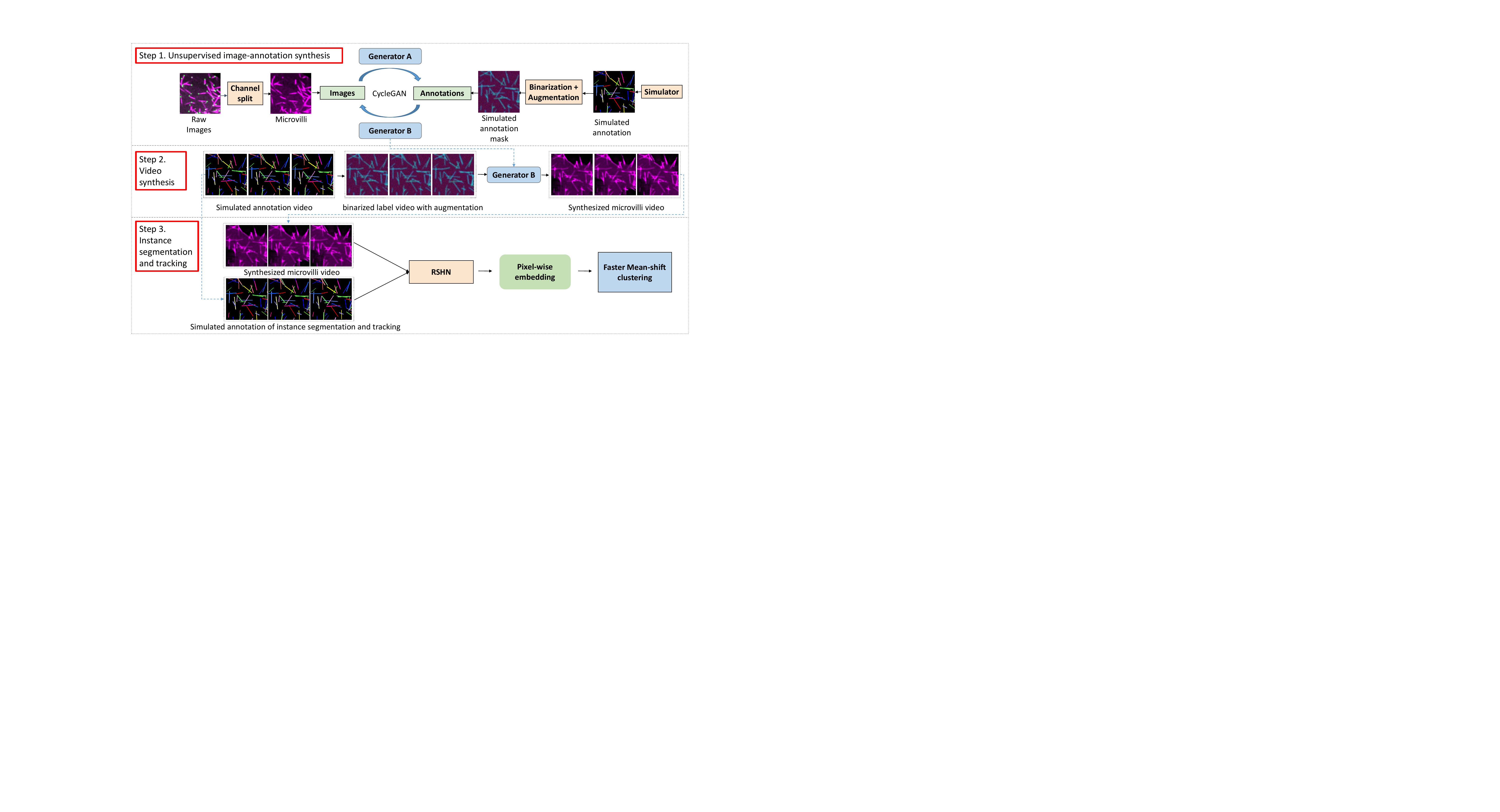}
\end{center}
   \caption{This figure shows the proposed ASIST method. First, CycleGAN based image-annotation synthesis is trained using real microvilli images and simulated annotations. Second, synthesized microvilli videos are generated by simulated annotation videos. Last, an embedding based instance segmentation and tracking algorithm is trained using synthetic training data.}
\label{fig:fig2}
\end{figure*}

As opposed to supervised learning, unsupervised learning aims to achieve quantitative measurements without using manual annotations. For example, a Cycle-GAN~\cite{zhu2017unpaired} based approach has been developed to perform annotation-free semantic segmentation on microvilli. However, few prior studies have achieve annotation-free instance segmentation and tracking on sub-cellular objects.

\section{Method}
\label{sec:Method}


In this paper, we propose a novel annotation-free synthetic instance segmentation and tracking (ASIST) method to characterize microscope videos containing sub-cellular objects. The annotation-free video quantification is enabled by generating synthesized microscope videos and annotations. With ASIST, the synthetic microvilli quantification is extended from semantic image segmentation to video analysis with instance segmentation and tracking (\textbf{Fig. \ref{fig:fig1}}).
The contribution of this paper is three-fold: (1) a new annotation-free video analysis paradigm is proposed. (2) The embedding based instance segmentation and tracking is aggregated with annotation-free synthetic learning as a holistic framework. (3) To the best of our knowledge, this is first study to investigate microvilli instance segmentation and tracking using embedding based deep learning.

The proposed ASIST method consists of three steps: (1) unsupervised image-annotation synthesis; (2) video synthesis with simulated instance  annotations; and (3) embedding based instance segmentation and tracking (Fig.\ref{fig:fig2}).

\subsection{Unsupervised image-annotation synthesis}
The first step of the ASIST method is to train a cycle-consistent generative adversarial network (CycleGAN)~\cite{zhu2017unpaired} to achieve image synthesis between microvilli images and simulated semantic labels. The training process, network design, and hyperparameters follows~\cite{liu2020gan}.

\subsection{Video synthesis}
We employ the segmentation-to-image generator (Generator B) from~\cite{liu2020gan} to generate synthetic microvilli video frames based on simulated annotation videos. The key task is to simulate instance annotations with consistent tracking numbers as a video. Briefly, we simulate $N$ instance labels in the first frame, following the same procedure as~\cite{liu2020gan}. Then, all instance labels in the first image frame are randomly translated, rotated and lengthened or shortened to form a simulated annotation video. Last, Generator B is used to synthesize the corresponding microvilli video in a frame by frame manner.

\noindent\textbf{Object number $N$}: The number $N$ can be flexibly defined. The values of $N$ are introduced in $\mathsection$\textbf{Experiment}.

\noindent\textbf{Translation}: The center coordinates of annotations are randomly translated by 1 pixel or stand still at 50$\%$ probability. 

\noindent\textbf{Rotation}: Each instance label are randomly rotated by 1 degree or stand still at 50$\%$ probability.

\noindent\textbf{Shortening/Lengthening}: Each instance randomly becomes longer/shorter by 1 pixels or be unchanged at 50$\%$ probability. One instance can only be longer/shorter in one video.

\noindent\textbf{Moved in/out}: Our video frames are generated in a larger size (e.g., 550 $\times$ 550 pixels) than targeting size (e.g., 512 $\times$ 512 pixels). Instances moving in and out are created by center cropping the larger size frame to the required size.

\subsection{Instance segmentation and tracking}
After obtaining paired synthetic microvilli and annotation videos, the recurrent stacked hourglass network (RSHN)~\cite{payer2018instance} is trained using the synthetic training data. Next, the pixel-wise feature embedding is achieved for the video, whose aim is to assign the same embedding values for all pixels belonging to the same object, while maximizing the differences of embedding between pixels from different objects. The Faster Mean-shift clustering algorithm~\cite{zhao2020faster} is used to achieve the final instance segmentation and tracking results.

\begin{figure}
\begin{center}
\includegraphics[width=0.8\linewidth]{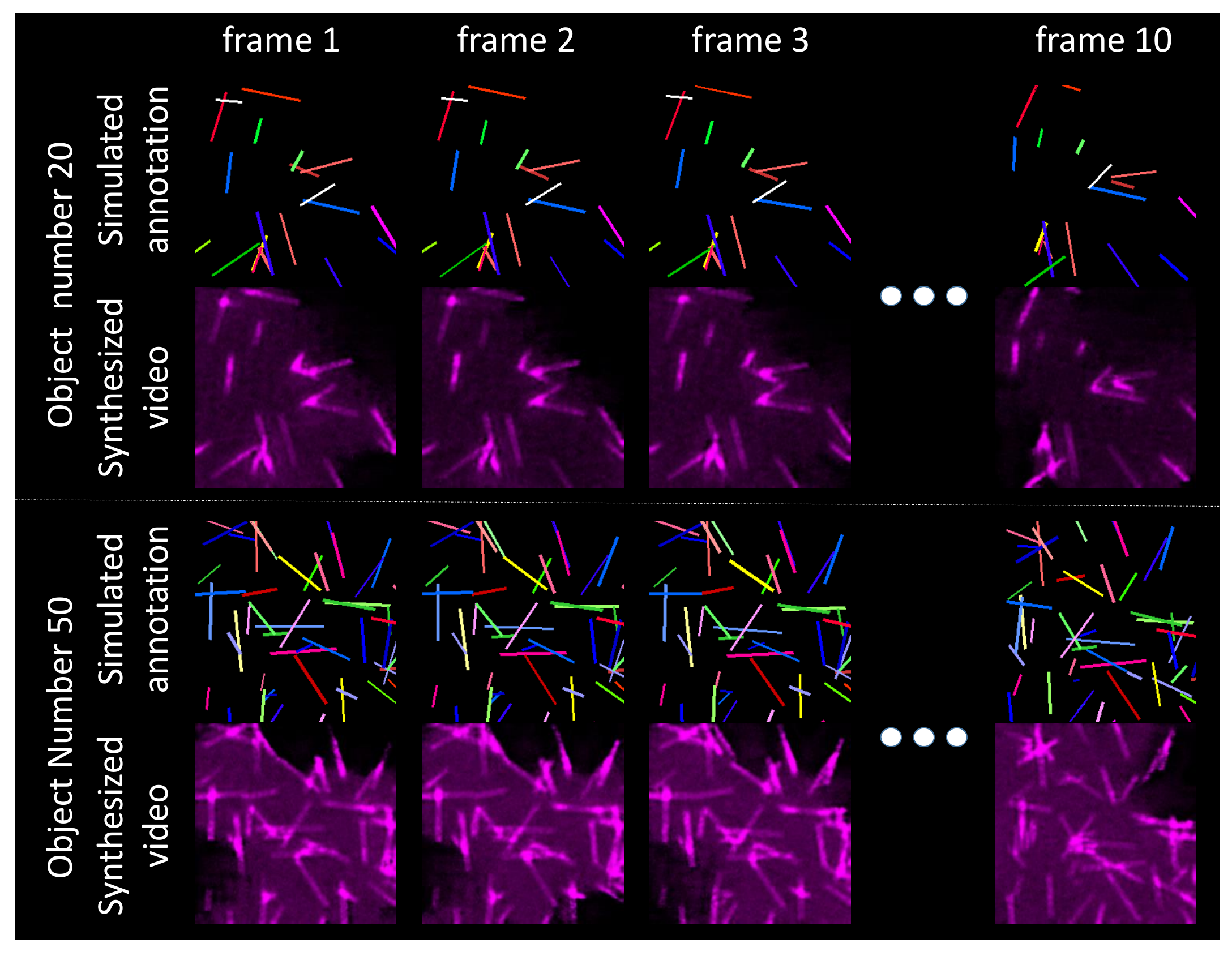}
\end{center}
   \caption{This figure shows synthesized microvilli videos and simulated label images. The different panels indicate the different number of simulated objects.}
\label{fig:fig3}
\end{figure}

\section{Experiments}
Two microvilli videos acquired using fluorescence microscopy were used as training and testing data respectively. Both videos had 1.1$\mu$m pixel resolution. One video with a resolution of 512$\times$512 pixels was used as training data, while the remaining videos with a resolution of 328$\times$238 pixels was used as testing data. The data was deidentified, and studies were approved by the Institutional Review Board (IRB). Limited by an extremely time-consuming manual annotation process, the first 10 frames of each video were annotated. Note that we not only annotated dense and highly overlapped objects with instance-level labels for each frame, but also assigned each instance a consistent tracking number across frames. It took one week of working hours for a graduate student to annotate two videos, which also demonstrates the need of annotation-free approaches for sub-cellular object quantification. 

In order to test our annotation-free instance segmentation and tracking model, we compared the performance of using the real microvilli video with manual annotations and using synthetic microvilli videos with simulated annotations. Briefly, following types of training data were employed:

\noindent\textbf{Upper} :the real testing microvilli video (328$\times$238) with manual annotation was used as training data, to compute the upper bound performance of using RSHN with 10 annotated frames.

\noindent\textbf{Real} : another real microvilli video (512$\times$512) with manual annotation was used as training data with 10 annotated frames.

\noindent\textbf{Simu-1}: 1 simulated video with 110 objects and 512$\times$512 image resolution, as decribted in § \textbf{Method}.

\noindent\textbf{Simu-5}: 5 simulated videos with 80, 110, 160, 200 and 220 objects and a 512$\times$512 image resolution.

\noindent\textbf{Simu-20}: 20 simulated videos by splitting each video in Simu-5 into four 256x256 videos (\textbf{Fig.\ref{fig:fig3}}). Herein, the total number of pixels of Simu-20 was the same as Simu-5.


\section{Results}
\begin{figure*}
\begin{center}
\includegraphics[width=0.91\linewidth]{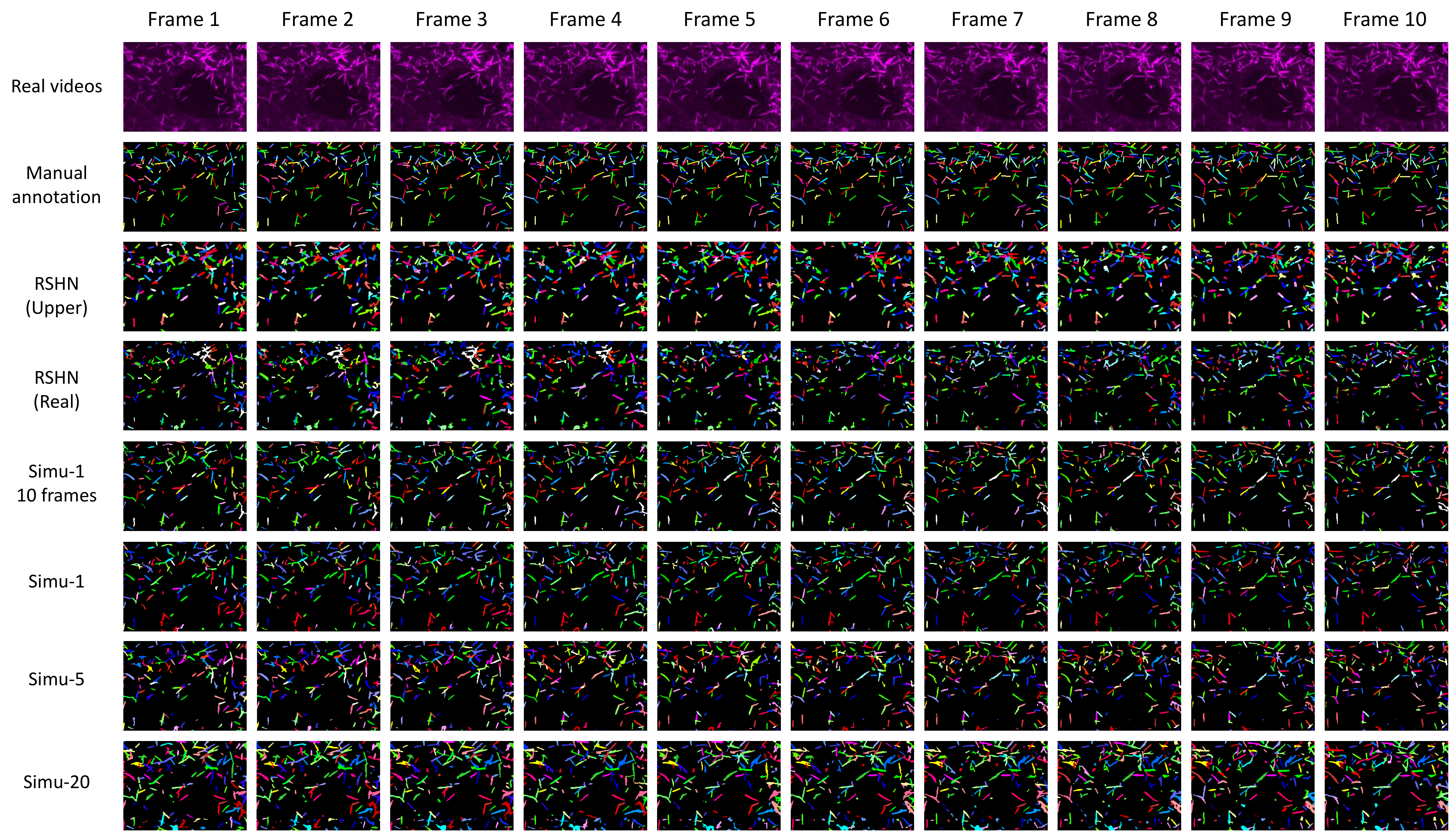}
\end{center}
   \caption{This figure shows the instance segmentation and tracking results of 10 frames of the testing real microvilli video.}
\label{fig:fig4}
 \end{figure*}

The qualitative result are shown in Fig.\ref{fig:fig4}. In our experiment, we used DET, SEG and TRA index computed by AOGM~\cite{matula2015cell} as our detection, segmentation, and tracking evaluation metrics respectively. Such metrics (the larger is the better) are the \textit{de facto} standard measurements in the Cell Tracking Challenge. The quantitative results of different methods are presented in Table.\ref{tab:TRA}. The metrics scores of using 20 synthetic videos achieved the best performance, which is better than using real data. Note that 2-3 days were spent to annotate one real microvilli video with 10 frames, while the synthetic videos did not require manual annotations for any length.

\begin{table}[]
\caption{DET, SET and TRA values of different experiments.}
\centering
\begin{tabular}{c|c|ccc}
\hline
Exp. & T.F. & DET & SEG & TRA \\
\hline
RSHN (Upper)~\cite{payer2018instance}& 10  & 0.6618 & 0.2983 & 0.6294 \\
\hline
RSHN (Real)~\cite{payer2018instance}& 10  & 0.3571 & 0.1687 & 0.3342 \\
ASIST (Simu-1) & 10 & 0.5798 & 0.3055 & 0.5508 \\
\hline
ASIST (Simu-1) & 50 & 0.5863 & 0.3109 & 0.5562 \\
ASIST (Simu-5) & 50 & 0.6601 & 0.3379 & 0.6268 \\
ASIST (Simu-20) & 50 & \textbf{0.7150} & \textbf{0.3316} &  \textbf{0.6735} \\
\hline
\end{tabular}
\noindent T.F. is the number of training frames of each video. RSHN (Upper) is the upper bound of RSHN of using testing video for training. RSHN (Real) is the standard testing accuracy using another independent video as training data.
\label{tab:TRA}
\end{table}

\section{Conclusion}
In this paper, we propose a new annotation-free framework, ASIST, to perform instance microvilli segmentation and tracking across microscopic videos without annotation. The annotation-free method achieved superior instance segmentation and tracking performance compared with the traditional supervised method due the convenience of simulating more heterogeneous and longer training videos, which is unsalable for manual annotations.


\bibliographystyle{IEEEbib}
\bibliography{refs}

\end{document}